\documentclass[pdflatex,sn-nature]{sn-jnl}

\usepackage{graphicx}%
\usepackage{multirow}%
\usepackage{amsmath,amssymb,amsfonts}%
\usepackage{amsthm}%
\usepackage{mathrsfs}%
\usepackage[title]{appendix}%
\usepackage{xcolor}%
\usepackage{textcomp}%
\usepackage{manyfoot}%
\usepackage{booktabs}%
\usepackage{algorithm}%
\usepackage{algorithmicx}%
\usepackage{algpseudocode}%
\usepackage{listings}%

\theoremstyle{thmstyleone}%
\theoremstyle{thmstyletwo}%

\theoremstyle{thmstylethree}%

\begin{document}

\title[Article Title]{Edge detection imaging by quasi-bound states in the continuum}


\author[1,2]{\fnm{Tingting} \sur{Liu}}

\author*[3]{\fnm{Jumin} \sur{Qiu}}\email{qiujumin@email.ncu.edu.cn}

\author[4]{\fnm{Lei} \sur{Xu}}

\author[5,3]{\fnm{Meibao} \sur{Qin}}

\author[3]{\fnm{Lipeng} \sur{Wan}}

\author[3]{\fnm{Tianbao} \sur{Yu}}

\author*[1]{\fnm{Qiegen} \sur{Liu}}\email{liuqiegen@ncu.edu.cn}

\author*[6]{\fnm{Lujun} \sur{Huang}}\email{ljhuang@phy.ecnu.edu.cn}

\author*[1,2]{\fnm{Shuyuan} \sur{Xiao}}\email{syxiao@ncu.edu.cn}

\affil[1]{\orgdiv{School of Information Engineering}, \orgname{Nanchang University}, \orgaddress{\city{Nanchang} \postcode{330031},  \country{China}}}

\affil[2]{\orgdiv{Institute for Advanced Study}, \orgname{Nanchang University}, \orgaddress{\city{Nanchang} \postcode{330031}, \country{China}}}

\affil[3]{\orgdiv{School of Physics and Materials Science}, \orgname{Nanchang University}, \orgaddress{\city{Nanchang} \postcode{330031}, \country{China}}}

\affil[4]{\orgdiv{Advanced Optics and Photonics Laboratory, Department of Engineering, School of Science and Technology}, \orgname{Nottingham Trent University}, \orgaddress{\city{Nottingham} \postcode{NG11 8NS}, \country{United Kingdom}}}

\affil[5]{\orgdiv{School of Education}, \orgname{Nanchang Institute of Science and Technology}, \orgaddress{\city{Nanchang} \postcode{330108}, \country{China}}}

\affil[6]{\orgdiv{School of Physics and Electronic Science}, \orgname{East China Normal University}, \orgaddress{\city{Shanghai} \postcode{200241}, \country{China}}}


\abstract{Optical metasurfaces have revolutionized analog computing and image processing at sub-wavelength scales with faster speed and lower power consumption. They typically involve spatial differentiation with engineered angular dispersion. Quasi-bound states in the continuum (quasi-BICs) have recently emerged as a powerful tool for tailoring properties of optical resonances. While quasi-BICs have been explored in various applications that require high $Q$-factors and enhanced field confinement, their full potential in image processing remains unexplored. Here, we demonstrate edge detection imaging by leveraging a quasi-BIC in an all-dielectric metasurface. This metasurface, composed of four nanodisks per unit cell, supports a polarization-independent quasi-BIC through structural perturbations, allowing simultaneously engineering $Q$-factor and angular dispersion. Importantly, we find that with suitable parameters, this quasi-BIC metasurface can perform isotropic two-dimensional spatial differentiation, which is the core element for realizing edge detection. Following the theoretical design, we fabricate the metasurfaces on the silicon-on-insulator platform and experimentally validate their capability of high-quality, efficient, and uniform edge detection imaging under different incident polarizations. Our results illuminate the mechanisms of edge detection with quasi-BIC metasurfaces and highlight new opportunities for their application in ultra-compact, low-power optical computing devices.}

\maketitle

\section{Introduction}\label{sec1}

The surge in global data and the real-time image processing demands of emerging technologies, such as augmented reality, biological engineering, and autonomous driving, have underscored the limitations of conventional digital circuit in computation speed and power consumption. Analog optical processing presents a viable solution to these challenges by enabling data manipulation at the speed of light and bypassing the need for analog-to-digital conversion, thus significantly reducing latency and energy consumption\cite{Solli2015}. Traditionally, analog optical computing leverages 4-f configurations for spatial filtering based on Fourier optics. While easy to implement, such systems with conventional lenses are inherently bulky and incompatible with compact integrated devices. Recent advancements in nanophotonics, particularly through the use of engineered metasurfaces, have provided an entirely new platform for realizing analog computing\cite{ZangenehNejad2020,He2022,Badloe2022,Neshev2023}. Metasurfaces have demonstrated unprecedented control of light propagation over a subwavelength thickness in beam steering\cite{Yu2011,Sun2012}, focusing\cite{Chen2018,Ou2022}, and holography\cite{Deng2018,Ren2020,Georgi2021}. They have been also developed in a wide range of mathematical operators by several theoretical and experimental studies\cite{Abdollahramezani2020,Silva2014,Kwon2018,Tanriover2023}. The core principle of computing metasurfaces is to engineer the nanostructures’ response to realize the optical transfer function of the desired operator in a small footprint. Consequently, there have been substantial efforts in the past few years to design metasurfaces for various analog operations for applications in imaging processing\cite{Wang2020b,Zhou2020,Wang2022}, beam shaping\cite{Divitt2019,Guo2020,Geromel2023}, and equation solving\cite{Cordaro2023}.

Edge detection based on analog differentiation operation is essential for rapid recognition and compression of object features in image processing. Metasurfaces have been proposed as compact analog differentiators using surface plasmon resonance\cite{Pors2014,Zhu2017}, spin Hall effect\cite{Zhu2019,He2024}, and Phanchartnam-Berry phase\cite{Zhou2019,Huo2020, Yang2024}. However, most existing work are restricted to one dimension or single-direction differentiation, and often require bulky optical components in the setup, which negates the advantages of metasurface compactness. An alternative approach involves designing metasurfaces with angle-selective optical response to achieve spatial differentiation by selectively filtering plane waves in momentum space without accessing Fourier space physically. The metasurface-based spatial differential operator, such as the Laplacian operator, are characterized by high-pass filters in Fourier space by transmitting waves at larger angles while suppressing those at smaller angles. In the context, high-index dielectric metasurfaces supporting leaky modes have demonstrated effective edge detection without resorting to the bulky 4-f lens systems, thereby reducing the footprint and enhancing integration\cite{Cordaro2019,Zhou2020a,Kwon2020,Ji2022,Cotrufo2023,Cotrufo2023a}. Recent advancements in dielectric metasurfaces have leveraged electric dipole resonance\cite{Wan2020}, magnetic dipole resonance\cite{Komar2021}, and toroidal dipole resonance\cite{Zhou2024}, to achieve desired angular responses. While the key to realizing edge detection is to design an optical resonance with required angular dispersion, a recently emerging concept called bound states in the continuum (BICs) provides abundant freedom of accurately controlling the spectral response (i.e., resonance position,  line width, angular dispersion) of optical resonances in a nonlocal metasurface\cite{Hsu2016,Huang2023}. For practical applications, it is necessary to convert BICs into quasi-BICs with finite and high $Q$-factors\cite{Koshelev2018,Wang2020,Huang2021}. Thanks to high $Q$-factors and extreme field confinement enabled by quasi-BICs, numerous interesting applications, including lasing\cite{Kodigala2017,Hwang2021,Ren2022}, sensing\cite{Yesilkoy2019,Luo2024}, vortex beam generation\cite{Wang2020a}, strong coupling\cite{Bogdanov2019,AlAni2021,Xie2024}, enhanced chirality\cite{Zhang2022,Chen2023,Shi2022}, and nonlinear effects\cite{Liu2019,Koshelev2020,Feng2023}, have been theoretically proposed and experimentally realized. Yet, edge detection based on a quasi-BIC metasurface remains unexplored. 

In this work, we experimentally demonstrate a dielectric metasurface for optical image edge detection by harnessing a polarization-independent quasi-BIC. We leverage a square lattice of silicon nanodisks supporting a quasi-BIC resonance to engineer its nonlocal response and achieve second-order spatial differentiation. By tailoring the spatial dispersion of the quasi-BIC resonance, the optical transfer function of the metasurface exhibits the feature of a sharp high-pass spatial filter corresponding to the Fourier transform of the Laplacian operation. Theoretical and experimental results confirm that the metasurface can produce an almost isotropic transfer function for Laplacian differentiation, enabling effective edge detection in nearly all directions. The high symmetry of the metasurface structure ensures polarization-independent operation. We emphasize that the silicon metasurface design allows to perform second-order spatial differentiation for edge detection without the need for any additional optical lensing and polarizing element, and the ultrathin silicon device, compatible with CMOS fabrication, thus can be directly integrated into existing imaging systems. This work presents a proof-of-principle implementation of a quasi-BIC metasurface-based image processing, which has potential applications in compact, low-power, and ultrafast all-optical data and image processing devices.

\section{Results}\label{sec2}

\subsection{General principle and metasurface design}

Our recipe to realize the 4f-less edge detection builds upon the general concept for two-dimensional (2D) Laplacian differentiation in a nonlocal metasurface and especially the free control of dispersive quasi-BIC resonances. Assuming a normally incident light beam along the $z$ axis, for example an optical image defined in the plane at $z=0$, with a transverse electric field profile $E_{\text{in}}(x,y)$, the input field can be decomposed into the a bundle of plane waves with different wave vectors following the spatial Fourier transform\cite{Deng2024,Cotrufo2024},
\begin{equation}
	E_{\text{in}}(x,y)=\int\int\tilde{E}_{\text{in}}(k_{x},k_{y})e^{i(k_{x}x+k_{y}y)}dk_{x}k_{y},
\end{equation}
where $k_{x}$ and $k_{y}$ are the wave vectors along the two orthogonal axes. For the purpose to realize second-order spatial differentiation, we would like to design the metasurface acting as optical filter on the transmitted light with electric field profile as $E_{\text{out}}(x,y)\propto\nabla^{2}E_{\text{in}}(x,y)$, where $\nabla^{2}$ is given by $\frac{\partial^{2}}{\partial x^{2}}+\frac{\partial^{2}}{\partial y^{2}}$. The task to realize Laplacian operation $\nabla^{2}$ on the light field in the real space is
\begin{equation}
	\nabla^{2}E_{\text{in}}(x,y)=-(k^{2}_{x}+k^{2}_{y})E_{\text{in}}(x,y)=-k^{2}_{\parallel}E_{\text{in}}(x,y).
\end{equation}
It can be observed that the metasurface for Laplacian operation should stastify the optical transfer function of $T(k_{\parallel})=-k^{2}_{\parallel}$, a quadratic functon of the in-plane wave vector $k_{\parallel}$. In Fourier space, the Laplacian operation can be described by
\begin{equation}
	E_{\text{out}}(k_{x},k_{y})=-(k^{2}_{x}+k^{2}_{y})E_{\text{in}}(k_{x},k_{y})=-k^{2}_{\parallel}E_{\text{in}}(k_{x},k_{y}).
\end{equation}
The optical transfer function in the wavevector space follows the function
\begin{eqnarray}
	t(k_{x},k_{y})&=&\begin{pmatrix}
		t_{ss}(k_{x},k_{y})  & t_{sp}(k_{x},k_{y}) \\ 
		t_{ps}(k_{x},k_{y})  & t_{pp}(k_{x},k_{y})  
	\end{pmatrix}\nonumber\\
	&=&\begin{pmatrix} 
		\alpha_{s}(k_{x}^{2}+k_{y}^{2})   & 0  \\ 
		0  & \alpha_{p}(k_{x}^{2}+k_{y}^{2}) 
	\end{pmatrix},\label{eq5}
\end{eqnarray}
where $s$ and $p$ on the first and second subscript represent the polarization of the incident and transmitted light, respectively. It is known that for an ideal Laplacian operator, the metasurface should have identical response to both polarizations $t_{ss}(k_{x},k_{y})=t_{pp}(k_{x},k_{y})\propto(k_{x}^{2}+k_{y}^{2})$ and have no polarization conversion $t_{ps}(k_{x},k_{y})=t_{sp}(k_{x},k_{y})=0$. But in many practical situations it is sufficient to achieve Laplacian differentiation with $\alpha_{s}\neq \alpha_{p}$\cite{Guo2018}. Because of the relationship of $k_{\parallel}=k\sin\theta$, where $k$ is the incident wave vector and $\theta$ is the incident angle. It is supposed that the metasurface exhibits transmission with zero value at normally incdience ($k_{\parallel}=0$ with $\theta=0$), and progressively increasing value for larger incident angles. Accordingly, the metasurface translates into a high-pass filter in Fourier space for the spatial differential operation. 

Next, we show how these stringent requirements for Laplacian operation can be met using a simple quasi-BIC dielectric metasurface. Figs. 1a-b schematically illustrate that the metasurface consisting of silicon nanodisks performs analog differentitation and transforms an image into its second-order derivative, leading to direct discrimination of the edges in the image. To realize the desired angle-dependent optical response of the metasurface, we use a unit cell comprising four silicon nanodisks sitting on the glass substrate. The structural parameters are denoted by the lattice constant $p$, the radii $r_{1}$ and $r_{2}$, and the thickness $h$ of nanodisks. Here we consider a metasurface operating in the near-infrared, while the operation window can be generalized to other spectral ranges. Within the wavelength regime of interest, we fix the period $p=1400$ nm, the radius $r_{1}=250$ nm, and thickness $h=220$ nm of nanodisks to meet the transmission requirement for differentation, and vary the raidius $r_{2}$ to tailor the quasi-BIC resonance. The design is numerically optimized using the finite difference time domain (FDTD) method. 

\begin{figure*}[!ht]%
\centering
\includegraphics[width=\textwidth]{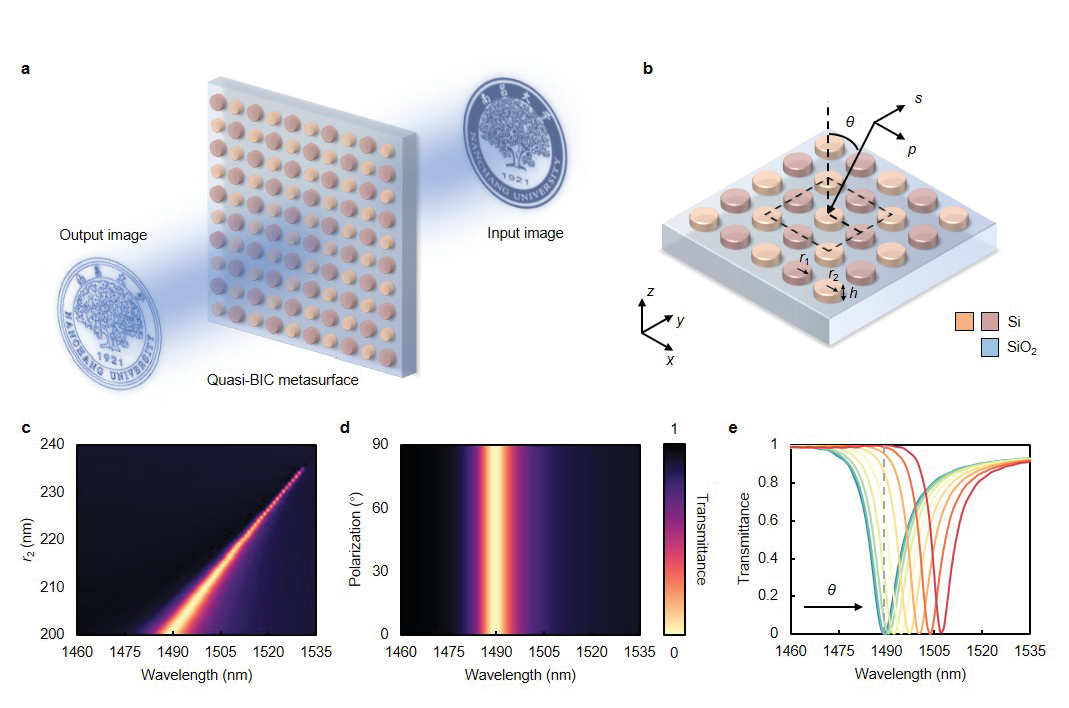}
\caption{\textbf{Quasi-BIC metasurface for edge detection imaging.} \textbf{a} Schematic of the proposed quasi-BIC metasurface which serves as a Laplacian operator to perform spatial filtering of images. \textbf{b} Unit cell of the metasurface composed of a square lattice of silicon nanodisks on a glass substrate.  \textbf{c} Simulated transmission spectra as functions of wavelength and nanodisk radius $r_{2}$. \textbf{d} Simulated transmission spectra under different polarization angles for $r_{2}=200$ nm.  \textbf{e} Simulated transmission spectra at different incident angles under $p$-polarization for $r_{2}=200$ nm. The incident angle is increased from 0$^{\circ}$ to 4$^{\circ}$ in 8 steps.}\label{fig1}
\end{figure*}

Unlike other resonance modes with frequencies below the light line, BICs exist as the embedded eigenstates within the continous spectrum of radiation modes in free space, completely decoupled from far-field radiation. By introducing pertubations to create specific radiation leakage channels, a genuine BIC transforms into a finite-lifetime leaky quasi-BIC resonance characterized by a Fano resonance in the reflection or transmission spectrum\cite{Koshelev2018,Xu2019,Wang2020,Zhou2022}. This evolution process can be observed in Fig. 1c, where a narrow transmission dip manifests energy leakage of system due to the perturbation from varying the radius $r_{2}$. As $r_{2}$ decreases from 250 nm, the resonance linewidth becomes broader, matching the prediction of reduced $Q$-factors of quasi-BICs. This suggests that the perturbations in BIC system would bring additional degrees of freedoms for engineering radiation and thus tailoring the resonance properties such as $Q$-factor and angular dispersion.

Given that the special arrangement of silicon nanodisks in the proposed metasurface exhibits $C_{4}$ rotational symmetry, it possesses polarization insensitivity. As illustrated in Fig. 1d, the transmission spectra of the quasi-BIC resonances remain almost unchanged by the incident polarization as the polarization angle varies from 0 to $\pi/2$, i.e., from $p$- to $s$-polarization. Furthermore, due to this symmetry, there is no polarization conversion output, thereby satisfying the zero conversion condition for the Laplacian operation. Under linear polarization excitation, the quasi-BIC resonance coupled to free space exhibits a Fano lineshape, with a rapid change in transmission as the incident angle increases. As an illustration, with $r_{2}=200$ nm under $p$-polarization illumination, as shown in Fig. 1e, the transmission spectral response of the metasurface demonstrates the desired angular dispersion behavior. At a wavelength of 1488.84 nm, the transmitted light gradually increases from 0 to 1 as the incident angle approaches $\theta_{\text{max}}=5^{\circ}$, which is a critical condition for realization of Laplacian operation. It is noteworthy that tailoring the quasi-BIC resonances by varying $r_{2}$ can broaden the transmission dip, allowing for an increase in $\theta_{\text{max}}$ and thereby expanding the detection bandwidth.

Following the above design, we investigate the angular dispersion of the metasurface under different polarizations. For the metasurface with nanodisks of radii $r_{1}=250$ nm and $r_{2}=200$ nm, the transmittances as functions of incident wavelength and angle are shown in Figs. 2a-c. Here, the transmittance for the unpolarized light is calculated as the average of the result for $p$- and $s$-polarizations. The symmetric arrangement of the nanodisks ensures that the resonances for different polarizations coincide at the wavelegnth of 1488.84 nm under normal incidence. At oblique incidence, the resonance wavelength in the transmission spectra display different degrees of redshift under different polarizations. Such an angle-dependent behavior preliminarily indicates that the metasurface can achieve spatial differentiation at the resonant wavelength across different polarizations.

\begin{figure*}[!ht]%
\centering
\includegraphics[width=\textwidth]{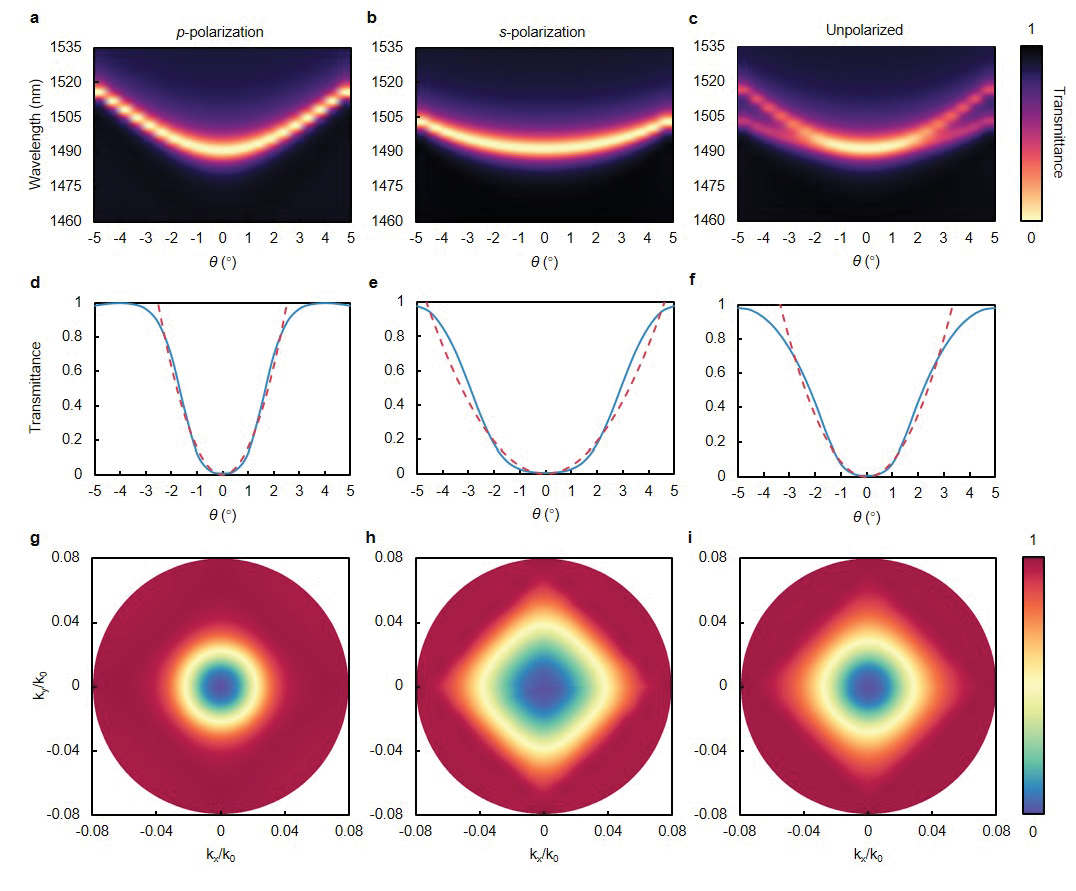}
\caption{\textbf{Simulated angular dispersion of the quasi-BIC metasurface with a nanodisk radius $r_{2}=200$ nm under different polarizations.} \textbf{a-c} Transmittance as functions of wavelength and incident angle. \textbf{d-f} Transmittance for different incident angles at the resonance wavelength 1488.84 nm and the corresponding quadratic fitting curves.  \textbf{g-i} Transmittance for different incident and azimuthal angles at 1488.84 nm.}\label{fig2}
\end{figure*}

To explore whether the transmittance is a quadratic function of the in-plane wave vector, we examine the transmittance versus the incidence angle at the resonant wavelength, as shown in Figs. 2d-f. Due to the strong dispersion of the leaky mode, the metasurface exhibits a rapid change in the transmittance as the incidence angle increases. We fit the transmission data for different polarizations with a quadratic function, representing the ideal second-derivative response, as indicated by the dotted lines. The transmittance shows good agreement with the ideal parabolic shape within the incidence angle up to $5^{\circ}$, leading to an NA of about 0.08 which corresponds to an edge resolution on the scale of 1 $\mu$m. This essentially determines the maximum spatial resolution for incident signals processed by the metasurface, as finer details correspond to larger transverse wavenumbers. Interestingly, the maximum angle as well as the resolution, NA, and the high-$k$ spectral feature, can be further improved by engineering the mode properties via the geometrical parameters of the metasurface. 

To further validate the differentation capability of the quasi-BIC metasurface, we numerically calculate the transmission at 1488.84 nm for arbitrary in-plane wave vectors and obtain the 2D optical transfer function across different polarizations in Figs. 2g-i. It is obvious that the dispesion response of the metasurface demonstrates high-pass filtering characteristics, suppressing low-frequency components and thereby producing sharp edges. The transmittances are nearly insenstive to the azimuthal angle and show almost circular contours for equal-transmission levels, suggesting the excellent isotropic property of the metasurface. Notably, the dispesion response adheres to the $C_{4}$ symmetry of the metasurface across different polarizations. The optical transfer funciton of the designed metasurface confirms its capablity to realize Laplacian operation for unpolarized waves. As the excitation of quasi-BIC resonances is general in dielectric metasurfaces, the optical transfer function can be realized at alternative wavelengths with different nanostructrues, enabling the application of differentation metasurfaces in various scenarios. 

\subsection{Experimental results}

Following the numerical simulations, we proceed with experimental demonstrations of the edge detection imaging. We fabricate silicon metasurfaces with structural parameters close to the numerically optimized design, using standard nanofabrication techniques, referring to the Methods section for detailed fabrication procedures. The scanning electron microscope (SEM) image of a sample is shown in Fig. 3a, with nanodisks of radii $r_{1}=250$ nm and $r_{2}=200$ nm. We employ a custom-built optical setup to characterize the optical response of the metasurface with respect to wavelength and incident angle. Fig. 3b illustrates that the measured transmission spectra at normal incidence features a Fano lineshape which can be fitted using the Fano formula $T=|a_{1}+ia_{2}+\frac{b}{\omega-\omega_{0}+i\gamma}|^{2}$, where $a_{1}$, $a_{2}$, and $b$ are real numbers, $\omega_{0}$ is the resonance frequency, and $\gamma$ is the leakage rate\cite{Miroshnichenko2010,Xu2019}. The transmission spectrum displays a dip around a wavelength of 1488.3 nm. The minimal transmission at normal incidence effectively suppresses low-frequency components of input optical images, thereby enhancing edge features. For the fabricated metasurfaces with different radii, the measured transmission spectra and their corresponding fittings are provided in Supplementary Information. The $Q$-factors defined by $\omega_{0}/2\gamma$ are extracted and presented in Fig. 3c. In the evolution process, enlarging the perturbation, manifested as the smaller radius $r_{2}$, gradually broadens the quasi-BIC resonances, indicating the additional degrees of freedoms in engineering radiation and thus the resonance properties such as $Q$-factor and angular dispersion. 

Fig. 3d displays the measured normal-incidence transmission spectra under different polarization angles ranging from 0 to $\pi/2$. As expected from the $C_{4}$ symmetry of the metasurface, the transmission spectra exhibits polarization-independent behaviors, showing a reasonably good consistency with simulations results in Fig. 1d. The angle dependent transmission measurements are also conducted for $p$- and $s$-polarizations, respectively. As shown in Figs. 3e-f, the quasi-BIC resonances with Fano lineshapes shift to longer wavelengths as the incident angle is increased. The measured spectra reasonably match the simulation results in terms of overall trends and shapes, showing significantly low transmission at normal incidence and a peak amplitude of near unity at the largest angle for the operating wavelength.

\begin{figure*}[!ht]%
\centering
\includegraphics[width=\textwidth]{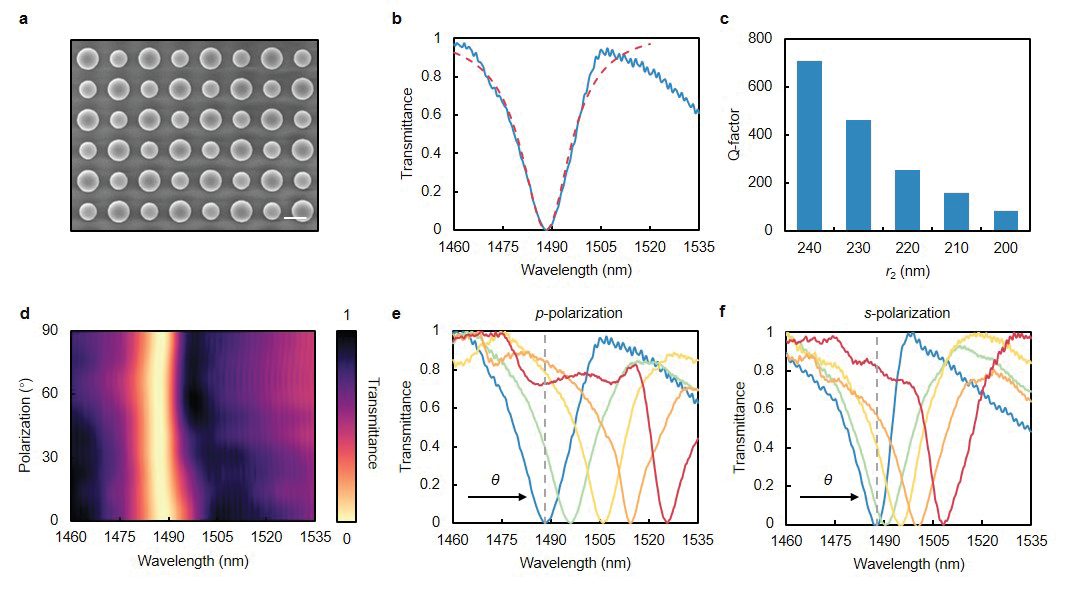}
\caption{\textbf{Experimentally measured optical response of the quasi-BIC metasurface.} \textbf{a} SEM image of the quasi-metasurface sample in the top view on the scale of 500 nm. \textbf{b} Measured transmission spectrum and the Fano fitting curve of the metasurface with a radius $r_{2}=200$ nm.  \textbf{c} $Q$-factors of the metasurface samples with varying radius $r_{2}$. \textbf{d} Transmission spectra of the metasurface with $r_{2}=200$ nm under normal incidence at different polarization angles.  \textbf{e,f} Transmission spectra for different incident angles under $p$- and $s$-polarizations. The incident angle is increased from 0$^{\circ}$ to 4$^{\circ}$ in 4 steps.}\label{fig3}
\end{figure*}

Next we demonstrate the edge detection capability of the proposed metasurface using the imaging setup shown in Fig. 4a, referring to the Methods section for detailed descriptions. The metasurface is used to detect the edges of a 1951 USAF resolution test chart. We select horizontal and vertical stripes, with stripe widths of 9.84 $\mu$m and equal spacing. We compare the imaging results without and with the differentiation operation of the metasurface in Figs. 4b-c. It is shown that the edges of these test images along the horizontal and vertical directions are clearly identified and recorded by the camera after passing through the metasurface. In our case, we experimentally investigate the influence of the incidence polarization on the edge detection. This is practically realized by inserting a half-wave plate after the light source in the experimental setup of Fig. 4a. The polarization of laser source is adjusted by rotating the half-wave plate, which has been experimentally verified through transmission spectra measurements. We record the output image for both the incident $p$- and $s$-polarizations. The edge detection is found to be almost identical under both polarizations, signifying polarization-independent and isotropic 2D spatial differentiation. We also plot the intensity distributions through the centers of the input and output images. It shows that the metasurface highlights the sharp changes of brightness in the images, with high-intensity peaks at the expected edge positions surrounded by nearly zero background. As a result, the edges become the most prominent part in the output. Additionally, each edge is accompanied by a pair of closely spaced, sharp peaks in the experimental measurements, further confirming the effectiveness and high quality of the metasurface in performing second-order spatial differentiation.

\begin{figure*}[!ht]%
\centering
\includegraphics[width=\textwidth]{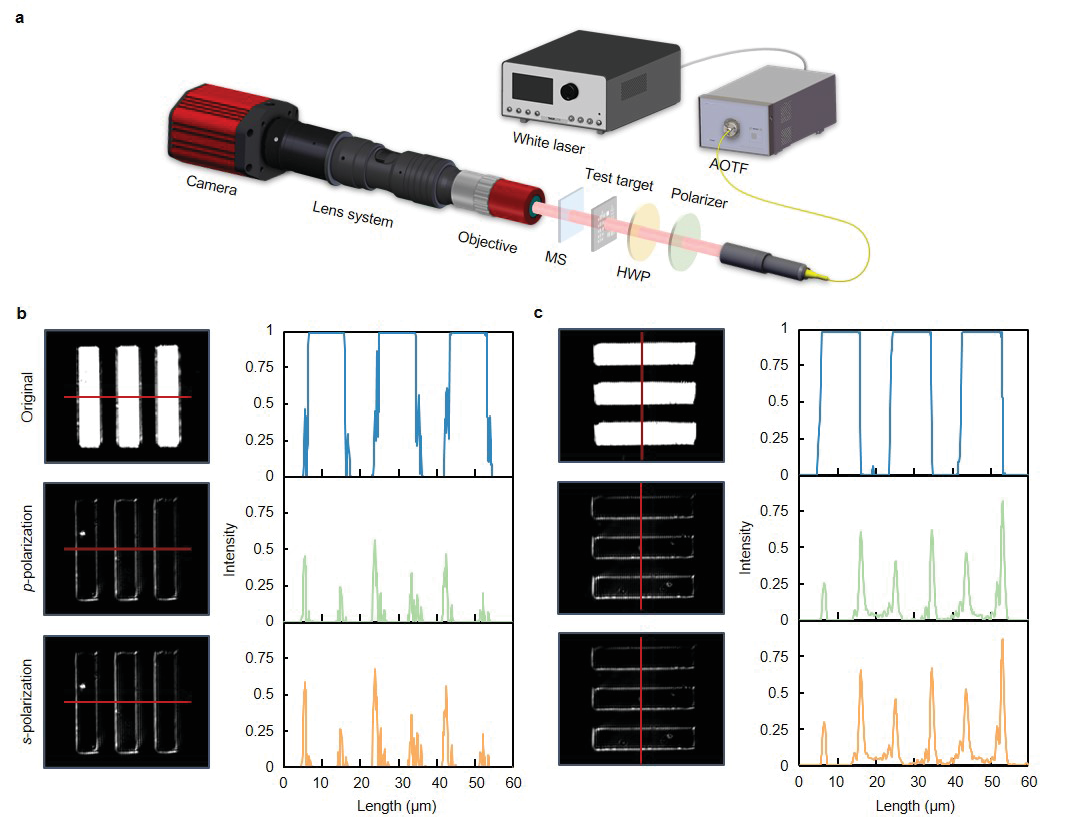}
\caption{\textbf{Edge detection imaging with the quasi-BIC metasurface.} \textbf{a} Schematic of the experimental setup. \textbf{b,c} Imaging results of the resolution test chart and the corresponding horizontal/vertical intensity profiles without and with the metasurface under different polarizations.}\label{fig4}
\end{figure*}

Finally, we prepare the input image with our institution logos, obtained by etching the desired shapes onto a chromium layer deposited on a glass substrate by electron beam lithography. The test images of these logos with the smallest width of approximately 2 $\mu$m (which approaches the resolution limit $\sim$ 1 $\mu$m of our metasurface) are shown in Fig. 5, and the sharp and high-contrast edges of both characters and lines can be clearly identified. This ensures the preservation of essential information while discarding redundant details, thus facilitating the image processing process. The quality of the edge detection and background suppression is further demonstrated by the horizontal intensity profiles provided in Supplementary Information. The ratio between the peak intensities of the output and input images approaches 1:5. These measurements confirm that the metasurface provides high-quality, uniform edge detection results. Moreover, in addition to the resonant wavelength, the experiments reveal an interesting phenomenon that the metasurface has a relatively broader working wavelength range. The edge detection imaging with the same metasurface at neighbouring wavelengths is also provided in Supplementary Information. These results can be attributed to the low-$Q$ of quasi-BIC resonance when $r_{2}=200$ nm. It is worth noting that this precisely highlights the merits of BICs physics. In contrast to other types of resonances, the $Q$-factor and the angular dispersion of the quasi-BIC resonances can be finely tuned by introducing varying degrees of perturbations to meet the practical requirements of optical transfer function.

\begin{figure*}[!ht]%
\centering
\includegraphics[width=\textwidth]{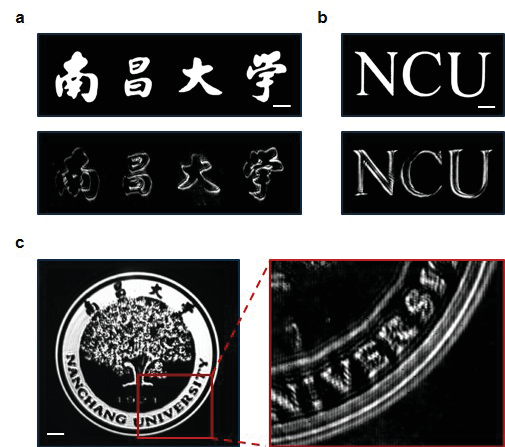}
\caption{\textbf{Experimental results of our institution logos with the quasi-BIC metasurface.} \textbf{a-c} Imaging and edge detection results for three types of Chinese characters, English abbreviation, and the badge of Nanchang University. Scale bar: 50 $\mu$m.}\label{fig5}
\end{figure*}

\section{Discussion}\label{sec13}

In conclusion, we have theoretically proposed and experimentally demonstrated an approach to design quasi-BIC metasurfaces for performing high-quality, isotropic, and polarization-independent 2D edge detection. This approach relies on precisely engineering the angular dispersion of the quasi-BIC resonance by fine-tuning the perturbation parameters in the nonlocal metasurface. Compared with previous metasurface differentiation operators relying on angular-selective transmission responses, the quasi-BIC metasurface offers enhanced flexibility and control over spectral features, allowing for the optimization of performance metrics in the spatial filtering process. We experimentally verify that a silicon metasurface characterized by Laplacian-like angle-selective transfer function arising from the quasi-BIC resonance, performs high-quality, efficient, and uniform second-order spatial differentiation under any incident polarization across a relatively broad wavelength regime. Our results suggest that the compact, low-power, and ultrafast optical edge detection can be implemented using quasi-BIC metasurfaces, showing great potential in various applications such as biological imaging and computer vision, especially when the BIC concept becomes general in the field of flat optics. Moreover, we anticipate that quasi-BIC metasurfaces with more complex dispersion engineering could also be realized by employing multilayer architectures and inverse design methods, leading to more sophisticated optical transfer functions and advanced functionalities in optical analog computing.

\section{Methods}\label{sec11}

\subsection{Numerical simulation}

The numerical simulations are performed using the FDTD method. The silicon nanodisks are positioned above a semi-infinite glass substrate. The periodic boundary conditions are applied to the sides of the square unit cell to model an infinite lattice. The perfectly matched layers on the top and bottom of the unit cell are used to absorb outgoing waves. The glass substrate and silicon are modeled as homogeneous and lossless materials, with a refractive index of 1.45 for the substrate and 3.4 for silicon across the wavelength of interest. To simulate the transmission response, the unit cell is illuminated by plane wave sources with the propagation oriented along the –$z$ direction, and a field monitor is used to capture the transmitted fields. The angular dispersion of the metasurface is calculated by varying the oblique incidence angles and the in-plane azimuthal angles.

\subsection{Sample fabrication}

The nanofabrication process of the metasurface samples begins with a silicon-on-insulator (SOI) wafer, featuring a top silicon layer of 220 nm and a buried layer of 2 $\mu$m, while the silicon substrate has a thickness of approximately 700 $\mu$m. Initially, a resist layer (ZEP520) is applied to the clean SOI wafer through spin-coating. Next, electron-beam lithography is used to create periodic patterns on the resist layer. These patterns are then etched into the silicon layer using an inductively coupled plasma method, with the resist acting as a masking layer. Finally, any remaining resist is removed using an N-methyl-2-pyrrolidone (NMP) solution. The fabrication was conducted at Tianjin H-Chip Technology Group Corporation. Scanning electron microscope (SEM) image of a metasurface sample is provided in Fig. 2a. 

\subsection{Optical Characterization}

For the transmission measurement, a supercontinuum white light laser (YSL Photonics, SC-Pro) is used as the incident light source for broadband wavelength scanning. The laser beam first passes through a linear polarizer, and then splits into two paths by a beam splitter, one of which is dedicated to optical path alignment using a visible camera (HIKVISION, MV-CS020-10UC), and the other is focused on the samples with a NIR objective (Mitutoyo, M Plan Apo NIR 20X / NA=0.40). The transmitted signal is finally collected by an optical spectrum analyzer (YOKOGAWA, AQ6370D) with an objective (OLYMPUS, UPlanFLN 20X / NA=0.50). The transmission measurement system is schematically depicted in Supplementary Information.

For the edge detection imaging, an acousto-optic tunable filter (YSL Photonics, AOTF-Pro) is employed to tune the wavelength to the center of the quasi-BIC resonance with a narrower band than the resonance spectral width. The test image, a 1951 USAF resolution test chart (LBTEK RB-N) is positioned in front of the sample. Through a magnification system consisting of a NIR objective (Mitutoyo, M Plan Apo NIR 50X / NA=0.42) and a tube lens (Navitar, 6.5X Zoom), the imaging results are captured by an infrared camera (First Light C-RED 2 Lite). The imaging system is schematically illustrated in Fig. 4a.

\section*{Data availability}

Data underlying the results presented in this paper may be obtained from the authors upon request.



\section*{Acknowledgments}

This work was supported by the National Natural Science Foundation of China (Grants No. 12364045, No. 12264028, and No. 12304420), the Natural Science Foundation of Jiangxi Province (Grants No. 20232BAB201040 and No. 20232BAB211025), the Chenguang Program of Shanghai Education Development Foundation and Shanghai Municipal Education Commission (Grant No. 21CGA55), the Shanghai Pujiang Program (Grant No. 22PJ1402900), and the Young Elite Scientists Sponsorship Program by JXAST (Grant No. 2023QT11).

\section*{Author Contributions}

T. L., J. Q., and S. X. conceived the original idea and the experiment. T. L., M. Q., and S. X. performed the design optimization. J. Q., L. X., and L. W. built the optical setup. J. Q. and S. X. conducted the optical measurements. T. L., J. Q., T. Y., and L. H. analyzed the data and prepared the figures. T. L. wrote the first draft of the manuscript. All authors contributed to finalizing the manuscript. Q. L., L. H., and S. X. supervised the project.

\section*{Competing Interests}

The authors declare no competing interests.

\end{document}